\documentclass[aps,prd,
onecolumn,groupedaddress,showpacs,nofootinbib,amssymb]
{revtex4-2}
\usepackage[dvips]{graphicx}
\usepackage{amssymb}
\usepackage{amsmath}
\usepackage{graphicx,color,comment,physics}
\usepackage{amsfonts}
\usepackage{bm}

\allowdisplaybreaks[4]

\begin{document}

\tolerance=5000

\newcommand{\matlab}{{\sc Matlab} }
\newcommand{\cvec}[1]{{\mathbf #1}}
\newcommand{\rvec}[1]{\vec{\mathbf #1}}
\newcommand{\ihat}{\hat{\textbf{\i}}}
\newcommand{\jhat}{\hat{\textbf{\j}}}
\newcommand{\khat}{\hat{\textbf{k}}}
\newcommand{\minor}{\mathrm{minor}}
\renewcommand{\trace}{\mathrm{trace}}
\newcommand{\spn}{\mathrm{Span}}
\newcommand{\rem}{\mathrm{rem}}
\newcommand{\ran}{\mathrm{range}}
\newcommand{\range}{\mathrm{range}}
\newcommand{\mdiv}{\mathrm{div}}
\newcommand{\proj}{\mathrm{proj}}
\newcommand{\R}{\mathbb{R}}
\newcommand{\N}{\mathbb{N}}
\newcommand{\Q}{\mathbb{Q}}
\newcommand{\Z}{\mathbb{Z}}

\newcommand{\e}{\mathrm{e}}

\title{Two Dimensional Conformal Field Theory \\
from Soft Quantum Electrodynamics in Four Dimensions}
\author{Shihab Fadda, Shin'ichi Nojiri$^{2,3}$}
\affiliation{
$^1$Department of Physics and Astronomy, Stony Brook University, Stony Brook, NY 11794-3800, United States\\
$^2$Department of Physics, Nagoya University, Nagoya 464-8602, Japan\\
$^3$Kobayashi-Maskawa Institute for the Origin of Particles and the
Universe, Nagoya University, Nagoya 464-8602, Japan
}


\begin{abstract}
In this work, we propose an effective action of the two-dimensional conformal field theory for the Soft modes appearing 
in Quantum ElectroDynamics (QED) in 4 dimensions. 
This is motivated in two ways. First, we motivate the notion of an effective action for the Soft modes using the path integral formalism. 
Second, by noting the transformations of the on-shell gauge-fixed QED action, we deduce necessary boundary conditions 
and the corresponding form of the counter-terms that govern the IR sector. 
We review a derivation of the Soft Ward-Takahashi identity using path integral methods and offer an alternative derivation 
that follows from the large-gauge variations of the QED action. 
Because our effective action is that of a conformal theory, we calculate the relevant central charge.  
\end{abstract}

\maketitle

\tableofcontents
\pagebreak
\section{Introduction}

The study of the low energy limit/behavior of physical models that describe Nature places important constraints on the parameters of the theory 
and dictates necessary regularization schemes that account for possible complications that arise. 
This is indeed the case for gauge theories described by a quantum theory of fields. 
For example, in the case of Quantum ElectroDynamics (QED), the re-summation of diagrams that include exchanges of an arbitrary number of low-energy 
virtual photons exponentiate and kill the amplitudes of all scattering processes. 
The suppression is logarithmic in the photon energy's infrared cutoff and arises due to the leading contribution to the Feynmann amplitudes from the addition of an external photon. 
This is the statement of Weinberg's Soft theorem \cite{Weinberg:1965nx}. 
The resolution to this issue, also due to Weinberg, is to realize that the Feynmann amplitudes are not physical, and neither is the assumption 
that we can resolve photons with arbitrarily small energies with our detectors. 
Indeed, the relevant scattering rates for processes, calculated by allowing a a finite amount of energy to go into undetected soft photons 
(defined as photons that possess energies that are below the sensitivities of our detectors), are rendered finite by explicit calculation. 
Very roughly speaking, while the individual amplitudes vanish, there are infinitely many amplitudes that contribute to these rates 
and cross sections, and thus we obtain a finite result. 
This a particular choice of the IR regularization scheme \cite{Weinberg:1995mt}.

This interesting interplay between the zeroes and the infinities has led many researchers to further investigate the origin of the leading order contributions to 
the Feynmann amplitudes, termed the leading Soft Theorem in the literature. 
An exciting breakthrough came when it was realized that Soft theorems and asymptotic symmetries are one and the same \cite{Strominger:2017zoo, He:2014cra}. 
The leading Soft terms are generated by the asymptotic symmetries of the gauge theory, and they can be related by studying modified versions of the Ward-Takahashi identities. 
The asymptotic symmetries in question, which are often elusive to define, are semi-local residual gauge symmetries that do not die-off at infinity (null infinity). 
Therefore, one can in principle define the associated Noether charges that generate these transformations. 
Perhaps surprisingly, these symmetries are spontaneously broken, as one can add an arbitrary number of ``Soft'' photons 
(now and hereon defined as photons with arbitrarily small energies) to any choice of vacuum and end up with an equally suitable vacuum. 
This line of thought has lead to significant progress in understanding Soft theorems, asymptotic symmetries and even holographic aspects of gauge theories \cite{Banerjee:2020zlg}.

An alternative, equally enlightening derivation of the leading Soft theorems utilizes the path integral formalism \cite{Avery:2015rga} 
to circumvent the necessity for asymptotic covariant quantization. 
In the path integral formalism, one can promote the semi-local symmetries, much like one does for global symmetries 
when deriving the conserved current, to obtain Ward identities directly from functional considerations. 
The form of the Ward identity is restricted by noting that the asymptotic symmetries are residual gauge modes, and thus leave the gauge-fixed action fixed. 
This is the approach that we will adopt and review for this work. 
In particular, we will use the technology developed \cite{Avery:2015rga} to obtain a streamlined derivation of the Soft Ward identities.

Having seen the utility of the path integral formalism first-hand, we wonder whether a holographic description of Soft theorems exists. 
We will see that this is indeed the case, and the form of such an effective holographic description can be inferred from considering boundary 
conditions in the path integration of non-interacting Maxwell theory. 
The resulting model; the Coulomb gas two dimensional 2D) conformal field theory (CFT) gives a dual description to many of the features of the Soft theorems, 
e.g., charge conservation, the Kac-Moody algebras, ... etc. 
We note that a similar result appeared in \cite{Kalyanapuram:2021bvf}, after an earlier version of this work was completed.

\section{Conventions}\label{Conventions}

\subsection{Space-time Conventions}

In this section, we introduce the various conventions that we use throughout the paper. This section closely follows the conventions of \cite{Strominger:2017zoo}. 
We start with the Minkowski space-time $(\mathcal{M},g)$, where $\mathcal{M}$ is a $1+3$ Lorentzian manifold and $g$ is 
the flat Minkowksi space-time metric, whose associated line element is:
\begin{equation}
\label{Mmetric}
ds^2=\sum_{\mu,\nu=0}^{3}\eta_{\mu\nu}\left(dx^{\mu}\right)\left(dx^{\nu}\right) \, ,
\end{equation}
where $\eta_{\mu\nu}$ is the usual symmetric and diagonal matrix whose diagonal elements are given by diag$\left(-1,1,1,1\right)$. 
If we now switch from the above co-ordinate system to a co-ordinate system given by the Bondi retarded co-ordinates, 
\begin{equation}
\label{future co-ordinates}
r=\sqrt{\sum_{i=1}^{3}\left(x^i\right)^2 } \, , \quad u=x^0-r\, , \quad \frac{2rz}{1+z\Bar{z}}=x^1+ix^2 \, , \quad r\frac{1-z\Bar{z}}{1+z\Bar{z}}=x^3 \, .
\end{equation}
Perhaps more usefully, we can rewrite the equation for $z$ as:
\begin{equation}
z=\frac{x^1+ix^2}{x^3+r}=\frac{\sin \theta}{1+\cos\theta}\mathrm{e}^{i\phi} \, ,
\end{equation}
where the second equality is written in the familiar spherical co-ordinates. 
Then the Minkowski space-time metric takes the following form, which is particularly useful near future null infinity, denoted $\mathcal{J}^+$,
\begin{equation}
\label{future metric}
ds^2=-du^2-2dudr+2r^2\gamma_{z\Bar{z}}dzd\Bar{z} \, ,
\end{equation}
where $\gamma_{AB}$ is the metric on the sphere of unit radius written in $z$ co-ordinates,
\begin{equation}
\label{spherical metric}
\gamma_{z\Bar{z}}=\gamma_{\Bar{z}z}=\frac{2}{\left(1+z\Bar{z}\right)^2} \, , \quad \gamma_{zz}=\gamma_{\Bar{z}\Bar{z}}=0 \, .
\end{equation}
A similar co-ordinate transformation (advanced Bondi co-ordinates) is useful when studying past null infinity, denoted $\mathcal{J}^-$,
\begin{equation}
\label{past co-ordinates}
r=\sqrt{\sum_{i=1}^{3}\left(x^i\right)^2 } \, , \quad v=x^0+r\, , \quad \frac{2rz}{1+z\Bar{z}}=x^1+ix^2 \, , \quad r\frac{1-z\Bar{z}}{1+z\Bar{z}}=x^3 \, .
\end{equation}
This gives the following flat space-time metric in advanced Bondi co-ordiantes,
\begin{equation}
\label{past metric}
ds^2=-dv^2+2dvdr+2r^2\gamma_{z\Bar{z}}dzd\Bar{z} \, .
\end{equation}
As we will be expanding various functions near future/past null infinity, we define the following expansion co-efficients for any arbitrary well-behaved function,
\begin{equation}
\label{r-expansion}
f\left(r,u,z,\Bar{z}\right)=\sum_{i}\frac{f^{\left(i\right)}\left(u,z,\Bar{z}\right)}{r^i} \, .
\end{equation}
The large-$r$ expansion will be useful when studying the QED E.o.M.s as well as the Cauchy data.

\subsection{Action and Gauge choice}

We start from the canonical action for QED,
\begin{equation}
\label{QEDAction}
S=S_{U(1)}+S_{\mathrm{M}}=-\frac{1}{4e^2}\int_{\mathcal{M}}F_{ab}F^{ab}+S_{\mathrm{M}}\, .
\end{equation}
Here $e$ is the electromagnetic coupling constant, $F_{ab} = \partial_a A_b - \partial_b A_a$, $S_{U(1)}$ is the action for the Maxwell field $A_a$ ($U(1)$ gauge field), 
and $S_{\mathrm{M}}$ is the action of matters. 
To ensure the Lorentz invariance of the action, the only interactions allowed are those that maintain the gauge symmetry, and thus are of the form,
\begin{equation}
\frac{\delta S_{\mathrm{M}}}{\delta A_a\left(x\right)}=- J^a\left(x\right) \, ,
\end{equation}
where in the above equation, $J^a\left(x\right)$ is proportional to the Noether current for a conserved global symmetry 
that is promoted to a local symmetry by the coupling to gauge fields. 
We therefore maintain the gauge arbitrariness, which we need to partially fix in order to solve the equations of motion (in covariant form), 
\begin{equation}
\label{maxwell}
\nabla_aF^{ab}=e^2J^b \, .
\end{equation}
We choose the following gauge and asymptotic condition in the retarded (advanced) Bondi co-ordinates,
\begin{equation}
\label{gauge}
A_r=0 \quad,\quad \lim_{r\rightarrow\infty}A_{u/v}=0 \, .
\end{equation}
We also expand the fields in inverse powers of the radius according to the following convention,
\begin{equation}
\label{large-r}
A_{\mu}\left(u,r,z,\Bar{z}\right)=\sum_m \frac{A_\mu^{(m)}\left(u,z,\Bar{z}\right)}{r^m} \, .
\end{equation}
Using the above gauge, the Euler-Lagrange equations of motion can be solved to give an asymptotic Laurent series of the gauge field-components in terms of the initial data,
\begin{equation}
\label{initial}
\left\{A_u^{(i)}\left(u=-\infty,z,\Bar{z}\right), A_z^{(m)}\left(u=-\infty,z,\Bar{z}\right), A_{\Bar{z}}^{(m)}\left(u=-\infty,z,\Bar{z}\right)\right\} 
\quad \mathrm{where} \quad i\in\{1,2,\dots\}, \, m\in\{0,1,2,\dots\}\, .
\end{equation}
For example, we have the following equations for the $z/\Bar{z}$ components of the gauge field \cite{Conde:2016csj},
\begin{align}
\label{cascade}
2\partial_uA_z^{(1)}&=\partial_zA_u^{(1)}+\partial_z\left[\gamma^{z\Bar{z}}\left(\partial_zA^{(0)}_{\Bar{z}}-\partial_{\Bar{z}}A^{(0)}_{{z}}\right)\right]+J_z^{(2)} \, , \\
2\partial_uA_z^{(m)}&=\left(1-m\right)A_z^{(m-1)}-\frac{2}{m}\partial_z\left[\gamma^{z\Bar{z}}\left(\partial_{\Bar{z}}A^{(m-1)}_{{z}}\right)\right]+J_z^{(m+1)} \, .
\end{align}
Since the recursion relations are those of the retarded time derivative, we can determine the above quantities in terms of the initial Cauchy data Eq.~(\ref{initial}).
If we now consider variations of the action about the solution in terms of arbitrary initial data modes, we find for the $S_{U(1)}$ term,
\begin{equation}
\label{variationS}
\delta S_{U(1)}=\frac{1}{e^2}\int_{\mathcal{M}}\delta A_a\nabla_bF^{ba}-\frac{1}{e^2}\int_{\partial\mathcal{M}}\delta A_a F^{ra} \, .
\end{equation}
The first term in the R.H.S. vanishes on-shell as it is just the Euler-Lagrange equation of motion (Maxwell's equation). 
The second term is usually ignored as we impose a phase-space variation that preserves the values of the field 
on the boundary (the Dirichlet boundary condition). 

We have seen that we have used some of the gauge arbitrariness in order to fix the gauge conditions Eq.~(\ref{gauge}). 
However, as is obvious by examining the remaining gauge degrees of freedom/integration constants, we still preserve the following residual gauge transformation,
\begin{equation}
\label{residual}
A_\mu\left(x\right)\rightarrow A_\mu\left(x\right)+\partial_\mu\epsilon\left(z,\bar{z}\right) \, .
\end{equation}
This residual gauge transformation is known as the large-gauge transformation, and plays an essential role in the study of the Soft Theorems. 
In section \ref{Large-Gauge Symmetry Breaking}, we show that while Eq.~(\ref{residual}) is conventionally a residual symmetry of the E.o.M., 
it gives rise to interesting variations of the action, that will guide us towards an effective Soft action.

\section{Large-Gauge Symmetry of the Action} \label{Large-Gauge Symmetry Breaking} 

\subsection{Non-Interacting Maxwell Theory}

To further investigate the role of the boundary terms in the study of the Soft Theorems and their holographic description, 
we turn to a study of the role they play in the context of local symmetries. 
The equations of motion are invariant under the large-gauge transformations, which correspond to a transformation of 
the initial data $\left\{A^{(0)}_z\left(u,z,\bar{z}\right)\, , \, A^{(0)}_{\bar{z}}\left(u,z,\bar{z}\right)\right\}$ as can be confirmed 
from the invariance of the first equation in Eq.~(\ref{cascade}) (as well as the gauge field's $u$-component equations which 
include $u$ and $r$ derivatives of the above Cauchy data, and are thus also invariant under the large-gauge transformations). 
As a simple example, we start by asking whether the free Maxwell action is invariant under Eq.~(\ref{residual}) at its minimum under the gauge conditions Eq.~(\ref{gauge}). 
After a straightforward calculation, we find that this variation is given by,
\begin{align}
\label{largegaugeaction}
\delta_\epsilon S_{U(1)}\left[\bar{A};\, \epsilon\left(z,\bar{z}\right)\right] =& 
 -\frac{1}{2e^2}\int_{S^2} d^2z \sqrt{\gamma}\, \left\{\nabla^B\epsilon\left(z,\Bar{z}\right)\left[A_B|^{u=\infty}_{u=-\infty}-A_B|^{v=\infty}_{v=-\infty}\right]\right\} \nonumber \\
=& -\frac{1}{2e^2}\int_{S^2} d^2z \sqrt{\gamma} \left\{ \nabla^B\epsilon\left(z,\Bar{z}\right)N_B \right\} \, .
\end{align}
In the above, barring indicates evaluating the fields of the action as those satisfying Eqs.~(\ref{maxwell}) and (\ref{gauge}) or equivalently Eq.~(\ref{cascade}), 
$B$ indices are to subsequently refer to unit sphere indices, raised and lowered by the appropriate metric on the unit sphere. 
The quantity $N_A$ is seemingly unconstrained by the E.o.M. as it is purely made of the Cauchy data.

We note that this is not an artifact of gauge fixing at the action level vs. gauge fixing at the E.o.M. level. 
Indeed, one has to be careful when fixing the gauge at the action level, as gauge-fixing and minimization 
of the action do not commute in general \cite{Motohashi:2016prk}. 
To avoid confusion regarding this point, we are NOT fixing the gauge at the action level prior to evaluating the equations of motion. 
We are simply re-writing the action via Stoke's theorem and evaluating it at the solution dictated by the Euler-Lagrange/Maxwell equations, 
solved with the gauge condition Eq.~(\ref{gauge}), with the equations invariant under the residual large-gauge transformations Eq.~(\ref{residual}).  
Instead of using Stoke's theorem first, suppose we wanted to find the variation by evaluating the action at the same solution and the Cauchy data 
phase-space configuration space in the arguments above, then taking the large-gauge variation. 
As one would can check after a swift calculation, and seemingly contrary to the result above, we have,
\begin{equation}
\label{actiongaugeinvariance}
\delta_\epsilon S_{U(1)}\left[\bar{A};\,\epsilon\left(z,\bar{z}\right)\right]=0\, .
\end{equation}
In fact, the above holds even for off-shell gauge configurations that satisfy the gauge conditions Eq.~(\ref{gauge}). 
Of course, this is what we expect initially and what our intuition told us would happen, the action is invariant under the residual gauge transformations.
To derive a more general result, we simply need to study the boundary terms that appear in the derivation of the Noether identity, also referred to as Noether's second theorem,
\begin{equation}
\delta S\left[A,\,\delta A\right]=\frac{1}{e^2}\int_{\mathcal{M}}\delta A_a\nabla_bF^{ba}-\frac{1}{e^2}\int_{\partial\mathcal{M}}\delta A_a F^{ra} \, .
\end{equation}
If we now study the special case of gauge-variations, and use the fact the action is gauge-invariant, we obtain,
\begin{align}
\label{noether2}
0=\delta_\epsilon S\left[A;\, \epsilon\left(x\right)\right]&
=\frac{1}{e^2}\int_{\mathcal{M}} \nabla_a\epsilon\nabla_bF^{ba}-\frac{1}{e^2}\int_{\partial\mathcal{M}}\nabla_a\epsilon F^{ra} \, ,\nonumber \\
&=-\frac{1}{e^2}\int_{\mathcal{M}} \epsilon\nabla_a\nabla_bF^{ba}-\frac{1}{e^2}\int_{\partial\mathcal{M}}\nabla_a\left(\epsilon F^{ra}\right) \,.
\end{align}
We evaluate the above to be,
\begin{equation}
\label{current?}
0=\delta S\left[\Tilde{A}; \,\epsilon\left(z,\bar{z}\right)\right]=-\frac{1}{e^2}\int_{\mathcal{J}^+}\nabla_u\left(\epsilon F^{ru}\right) 
 -\frac{1}{e^2}\int_{\mathcal{J}^-}\nabla_v\left(\epsilon F^{rv}\right) \, .
\end{equation}
The quantities above are off-shell, with the tilde indicating that the fields satisfy the gauge choice, and in going from the last line of Eq.~(\ref{noether2}) 
to Eq.~(\ref{current?}), we have used the fact that the total divergence includes a spherical divergence, which vanishes due to the sphere having no boundary. 
We take the residual gauge transformations in order to avoid variations that result from the gauge-fixing functionals. 
The bulk integral is the well-known Noether identity, which vanishes identically, even off-shell,
\begin{equation}
\label{noether}
\nabla_a\nabla_bF^{ab}=0 \, .
\end{equation}
The above Noether identity is interpreted as the linear-dependence, up to some integration constants, among the E.o.M. due to the gauge redundancy. 
Eq.~(\ref{current?}) tells us that we may define a current,
\begin{equation}
\label{conservedcurrent}
 j^b=\frac{1}{e^2}\nabla_a\left[\epsilon\left(z,\bar{z}\right)F^{ab}\right]\, ,
\end{equation}
which, even when off-shell, satisfies the conservation,
\begin{equation}
\nabla_bj^b=0\, .
\end{equation}
In fact, one could define a conserved current for a local-gauge transformation much like the above, with a general space-time co-ordinate replacing the co-ordinates on a 2-sphere. 
However, that conservation of such a current would require that the gauge degrees of freedom are un-fixed and its insertions within the path-integral formalism 
[or equivalently, its quantization] will thus be ill-defined. 
Therefore, such a construction can only be done for the large-gauge degrees of freedom in some sense. 
The above conservation equation is related to Noether's identity Eq.~(\ref{noether}) much like global symmetry currents utilize the E.o.M. to satisfy conservation on-shell. 
The conservation of this current leads to the boundary large-gauge variation Eq.~(\ref{largegaugeaction}). 
Adding to that resemblance, one can define a conserved ``Soft'' charge on a space (or null)-like hyper-surface; $\Sigma$,
\begin{equation}
\label{softcharge}
Q\left[\epsilon\left(z,\bar{z}\right)\right]=\int_{\Sigma} d\Sigma_aj^a\,.
\end{equation}
We note that this is identical to the charge defined in \cite{Strominger:2017zoo} even the interacting setting 
(with the assumption that the radiative modes vanish near time-like infinities). 
The above definition is valid even for interacting theories, since it only uses the properties of the free Maxwell strength tensor, 
and is thus always conserved. 
Additionally, it was shown in \cite{He:2014cra} that it generates the large-gauge transformations. 
An important distinction is that that the charge in the literature is associated to co-dimension $2$ surface, while the above is that of a co-dimension $1$ surface. 
The subtlety lies in whether one should consider the boundary at time-like infinity and its contributions. 
These contributions were assumed to vanish in the references, but play a critical role in the theory; they define the vacuum. 
We elucidate this point further in the following sub-sections.

To see how the Eqs.~(\ref{noether2}) and (\ref{current?}) are ultimately equivalent to Eq.~(\ref{largegaugeaction}), 
one needs to go on-shell by employing the Gauss constraint on the Cauchy initial/final surfaces; past/future null infinity,
\begin{equation}
\label{gauss}
0=\left(dr\right)_b\nabla_aF^{ab} \, .
\end{equation}
On-shell, one may use the E.o.M. (with no matter coupling) and the gauge conditions Eq.~(\ref{gauge}) to find Eq.~(\ref{largegaugeaction}) up to a factor of $2$, 
which is immaterial as the integral vanishes identically (one can always take linear combinations of Eq.~(\ref{largegaugeaction}) and Eq.~(\ref{actiongaugeinvariance}) 
to obtain any constant in front of the boundary term). 

The luxury afforded to us by the absence of matter allows us to show that the action is invariant under large gauge transformations, 
as can be seen from the equality of these two expressions. 
Once we couple the gauge fields to matter, we see that we must account for the $U(1)$ global symmetry current in the E.o.M. and therefore the variation 
of the action under the large-gauge symmetry on-shell is non-zero in general.

\subsection{Scalar QED}

In this section, we consider the general form of the conserved currents that result from interacting theories. 
We will be handling the case of massless scalar QED for illustration and concreteness, but the general features are generalizeable to any QED (or QCD). 

Consider the action given by,
\begin{equation}
\label{QED}
S_{\mathrm{QED}}=-\frac{1}{4e^2}\int_{\mathcal{M}}F_{ab}F^{ab}+\int_{\mathcal{M}}\left(D_a\phi\right)^* D^a\phi+m^2\left|\phi\right|^2\, ,
\end{equation}
where $D_a=\nabla_a-ieA_a$ is the covariant derivative. 
Taking the variation, and restricting ourselves to the case of pure gauge variations of both the gauge field and the complex scalars, we obtain,
\begin{align}
\delta_{\epsilon}S_{\mathrm{QED}}\left[A,\,\phi,\,\phi^*;\,\epsilon\right]
=&\int_{\mathcal{M}}\epsilon\left[-\frac{1}{e^2}\nabla_a\nabla_bF^{ab}-\nabla_aJ^a\right]+i\epsilon e\left[D_aD^a\phi-\left(D_aD^a\phi\right)^*\right] \nonumber \\
&+\int_{\partial\mathcal{M}}\left[\frac{-1}{e^2}\nabla_a\left(\epsilon F^{ra}\right)+i\epsilon J^r\right]-i\epsilon e\left[D^r\phi-\left(D^r\phi\right)^*\right]\,.
\end{align}
If we now go on-shell (in particular the $U(1)$ current contribution; $J^a$ is eliminated in the bulk and on the boundary),  
and again impose the gauge and fall-off conditions Eq.~(\ref{gauge}), restricting our currents to those of the large-gauge modes, we obtain,
\begin{equation}
\label{interacting}
\delta_{\epsilon}S_{\mathrm{QED}}\left[\bar{A},\,\bar{\phi},\,\bar{\phi^*};\,\epsilon\right]
=\int_{\partial\mathcal{M}}-\frac{1}{e^2}\epsilon \nabla_BF^{rB}-ie\epsilon\left[D^r\phi-\left(D^r\phi\right)^*\right]\, .
\end{equation}
In the above, we used integration by parts (on the 2-sphere) to showcase the explicit dependence on $\epsilon$ instead of its derivatives, 
as we will be needing that when we study the Ward identities. 
It is clear that we must now turn our attention to the third term, arising from the complex scalar equations of motion as a potential new player in the game. 
For this particular case, we refer to \cite{Campiglia:2015qka} and borrow the result for near null-infinity regions,
\begin{equation}
\label{nullregion}
\phi\left(x\right)\sim O\left(r^{-\frac{3}{2}}\right)\, .
\end{equation}
This is all we need to show that the extra term actually vanishes for on-shell solutions with appropriate asymptotic behavior. 
Therefore, we retrieve back our original result Eq.~(\ref{largegaugeaction}),
\begin{equation}
\label{idkw}
\delta_{\epsilon}S_{\mathrm{QED}}\left[\bar{A},\,\bar{\phi},\,\bar{\phi^*};\,\epsilon\right]
= - \int_{\partial\mathcal{M}} \frac{1}{e^2}\epsilon \nabla_BF^{rB}\, .
\end{equation}
However, unlike the case with the free Maxwell theory, there is no trivially conserved current which forces the variation of the on-shell action 
under the large-gauge variation of our scalar QED action to vanish. 
This is a slight problem. 
To see why, recall that the well-defined variational prescription enforces certain boundary conditions on our dynamical variables. 
Traditionally, one can impose the Dirichlet boundary condition if one controls or has a way of knowing the Cauchy data. 
We are certainly not in this situation. 
Indeed, there is no physical way to tell apart two gauge fields that differ by a gauge transformation. 
The gauge degrees of freedom are, for all intents and purposes (apart from the Lorentz invariance purposes), redundancies in our description; 
they should not manifest themselves as a non-trivial variation of the action at its supposed stationary point, spoiling its stationary. 
The boundary conditions must then set factors multiplying $\epsilon$ to identically vanish,
\begin{equation}
\label{bc}
\mbox{boundary condition} \Rightarrow \frac{1}{e^2}\nabla_BF^{rB}|_{\partial\mathcal{M}}=0\, .
\end{equation}
This motivates the introduction of an IR counter-term, serving as an effective Soft action on $\partial\mathcal{M}$ which enforces this boundary condition. 
Thus, we have all the necessary ingredients to discuss our main result in Section 4. Before we do so, we discuss how this is related to the Soft theorem. 

Boundary conditions play an important role in path-integration, despite being irrelevant to the every-day Feynmann-diagrammatic calculations 
which are mainly interested in the bulk and the higher-than-IR behavior of a theory. 
If one ignores the boundary conditions for a second, we can evaluate the variation of a string of operators within a path integral under the following field re-definitions,
\begin{align}
A_a\left(x\right)&\rightarrow A_a\left(x\right)+\partial_a \epsilon\left(z,\bar{z}\right) \, ,\nonumber \\
\phi\left(x\right)&\rightarrow \e^{iq\epsilon\left(z,\bar{z}\right)}\phi\left(x\right)\,.
\end{align}
Then the variation of the path integral with a string of operators; $X$, consisting both of gauge and matter fields and denoted as vacuum correlation function, becomes,
\begin{align}
\label{Ward}
0=\delta_{\epsilon}&\left<\mathrm{VAC, Out}\right|X\left|\mathrm{VAC, In}\right>
=-i\int_{\mathcal{J}^+}\frac{1}{e^2}\epsilon\left<\mathrm{VAC, Out}\right| \nabla_BF^{rB}X\left|\mathrm{VAC, In}\right> \nonumber \\
&-i\int_{\mathcal{J}^-}\frac{1}{e^2}\epsilon\left<\mathrm{VAC, Out}\right|X\nabla_BF^{rB}\left|\mathrm{VAC, In}\right>
+\left<\mathrm{VAC, Out}\right|\delta_{\epsilon}X\left|\mathrm{VAC, In}\right>\,.
\end{align}
This is the exact result of the Soft Theorem Ward identity obtained in \cite{Avery:2015rga}, obtained here via straightforward analysis of the action. 
However, the keen reader might notice a dangerous contradiction. 
Namely, we have assumed that our ``vacuum'' is invariant under large-gauge transformations. 
If one defines conserved charges that generate the large-gauge transformations, then we immediately have a contradiction, 
by noting that large-gauge transformations act homogeneously on gauge fields and thus provide a classical case of spontaneous symmetry breaking. 
Indeed, we have seen a trivially conserved charge that satisfies the above conditions in Eq.~(\ref{softcharge}) and the comments below. 
This would suggest that we must add a vacuum shift term contributing an incorrect factor of  $2$ to our charge. 
Indeed, this is quite unusual, since we expect our states to be independent of the gauge choice seeing that we are interested in physical states, i.e., 
the ones lying in the cohomology of the BRST operator. 
For physical states the above result thus should hold true. 
However, in the path-integral prescription, the in/out states are functionals given in terms of the set of fields at time-like infinity, 
whose dependence on gauge parameters may render the state-functional not invariant under large-gauge transformations. 
Such is the case with the traditionally defined of the vacuum functional. 
This begs the following questions:  
What is our Hilbert space? 
How can a gauge transformation manifest in a physical phenomenon, such as the IR behavior of a theory? 
This is often a point that causes a lot of confusion, and that should be addressed via a more rigorous treatment. 
The authors are unaware if such treatment has been done. 
The derivation in the following sub-section, following the arguments in \cite{Avery:2015rga}, circumvents this problem, by circumventing time-like infinities altogether. 

\subsection{Alternative Derivation} \label{ad}

To illustrate the essential aspects of the derivation, we consider the Schwinger-Dyson equation for the QED action Eq.~(\ref{QEDAction}), 
in the presence of a string of operators; $X$,
\begin{equation}
\label{X}
X:=A_{a_1}\dots A_{a_n}\Psi_{i_1}\dots\Psi_{i_m}\,.
\end{equation}
The gauge fixing functional will be denoted,
\begin{equation}
\label{GF}
G=B\left[f\left[A\right]\right]\mathrm{Det}\left[\mathcal{F}\right]\, .
\end{equation}
We now consider the Schwinger-Dyson equation of the gauge fields, found by varying the correlators under an arbitrary compact vector field $\epsilon_\mu\left(x\right)$,
\begin{equation}
\label{vA}
A_\mu\left(x\right)\rightarrow A_\mu\left(x\right)+\epsilon_\mu\left(x\right)\,.
\end{equation}
We find,
\begin{align}
\label{WI1}
0=i\int_{\mathcal{M}}\epsilon_\nu\nabla_\mu\left<F^{\mu\nu}X\right>
+\int_{\mathcal{M}}\epsilon_\mu\left<\frac{\delta G}{\delta \epsilon_\mu}X\right>-i\int_{\mathcal{M}}\epsilon_\mu\left<J^\mu X\right>+\left<\delta_\epsilon X\right>\,.
\end{align}
In the above, the second term corresponds to a gauge choice change. 
Note that we are explicitly considering field insertions that are explicitly non-invariant under the gauge transformations, 
therefore the variation of the gauge-fixing functional has to be taken into account in general. 
In the special case where the variation is pure gauge over a compact region, or more precisely,
\begin{equation}
\label{WI2}
\epsilon_\mu\left(x\right)=\rho_R\left(x\right)\nabla_\mu\Omega\,,
\end{equation}
where $\rho_R\left(x\right)$ is the indicator function of a compact region; $R$, defined to satisfy,
\begin{equation}
\label{WI3}
\rho_R\left(x\right)=
\begin{cases}
1& \mbox{if } x\in R\\
0& \mbox{otherwise}
\end{cases}
\end{equation}
As we are explicitly working with the assumption that the vacuum is spontaneously broken \cite{Matsumoto:1973hg}, 
we require $R$ not to include time-like infinities as to avoid the region of definition of functionals of asymptotic states (including the vacuum functional). 
Due to $R$ being a compact region, we can integrate by parts over the manifold for free. 
Thus we get,
\begin{align}
0=& -i\int_{R}\nabla_\nu\nabla_\mu\Omega\left<F^{\mu\nu}X\right>+\int_{\mathcal{M}}\left<\delta_{\rho,\Omega} GX\right>
+i\int_{R}\Omega\nabla_\mu\left<J^\mu X\right>+\left<\delta_{\rho, \Omega} X\right>\nonumber \\
& +\int_{\partial R}n_{\mu}\left[\nabla_\nu\Omega\left<F^{\mu\nu}X\right>-\Omega\left<J^{\mu}X\right>\right]\,.
\end{align}
We are interested in the transformations that leave the gauge-fixing functional invariant; the residual gauge transformations. 
For the choice of gauge Eq.~(\ref{gauge}), the indicator function poses no alteration to the gauge-fixing form, and thus is invariant in the above equation. 
A more careful analysis would be needed for gauge-conditions involving derivatives, such as the Coulomb or Lorentz gauge conditions. 
For the large-gauge transformations, we take the region $R$ to include all the points appearing in the string $X$ and use the Ward-Takahashi identities 
for the QED conserved current to obtain,
\begin{equation}
\label{LSZ}
0=\left<\Tilde{\delta}_{\rho, \Omega} X\right>+\int_{\partial R}n_{\mu}\left[\nabla_A\Omega\left(x^A\right)\left<F^{\mu A}X\right>-\Omega\left<J^{\mu}X\right>\right]\,.
\end{equation}
In the above, $\Tilde{\delta}$ denotes the full large-gauge transformation, including that of both gauge and matter fields, the latter's generated by the Ward-Takahashi identity. 
By the arguments given in \cite{Avery:2015rga}, we can then push our compact region $R$ to be arbitrarily close to the ``boundary'' of our space-time. 
We assume the vacuum has no matter degeneracy and thus we can ignore the last term, and we interpret the insertion of the electromagnetic strength tensor 
at the boundary as a Soft photon insertion. 
This result is identical to Eq.~(\ref{Ward}). 
The LSZ prescription then gives the usual Soft Theorem. 

We now explicitly show how the vacuum shift arises from the above assumptions. 
This is most easily seen if one takes $X=A_B\left(x\right)$ into Eq.~(\ref{LSZ}), which by Eq.~(\ref{softcharge}) 
and the assumption of no matter degeneracies is equivalent to the following commutator,
\begin{equation}
\label{commutator}
\left<\mathrm{VAC, Out}\right|\left[Q_\epsilon\, , \, A_B\left(x\right)\right]\left|\mathrm{VAC, IN}\right>
=i\partial_B\epsilon\left(x\right)\left<\mathrm{VAC, Out}|\mathrm{VAC, In}\right>\, .
\end{equation}
But famously, the above indicates that the under non-trivial representations of the symmetry group. 
Therefore, this indicates that our local symmetry is spontaneously broken. 
We discuss this matter further in Appendix \ref{A1}. 
We are now ready to discuss the effective IR Action/counter-term.

\section{Soft Effective Action}

\subsection{Path Integral Argument}

In this subsection, we offer a path integral argument, similar to the ones made when discussing holography, to motivate the use 
of boundary terms of an action as an effective Soft action. 
The arguments made in this subsection rely on schematics are therefore are not to be taken literally. 
Nevertheless, we feel that they may provide an exciting intuitive introduction to the notion of an effective action on the boundary. 
Consider the following familiar path integral for free four dimensional QED, 
\begin{equation}
\label{SP1}
D_\mathrm{QED}= \int_{-\infty}^{\infty}\prod_{\mu,x}\left( 
dA_{\mu}\left( x \right)\right)\det[\mathcal{F}]B\left[g\left[A_\mu\right] \right] 
\exp \left[-\frac{i}{4e^2}\int_{\mathcal{M}} F_{\mu\nu}F^{\mu\nu}+i\epsilon \, \mathrm{terms} \right] \, .
\end{equation}
The field eigenvalues, $A_{\mu}\left( x\right)$ are those of the photon gauge fields 
in the canonical theory. 
Here, $\mu$ and $\nu$ run over the Lorentz labels $\{t,x,y,z\}$, and $x$ is representative of 
the Lorentzian  coordinate, $\left(t,x,y,z\right)$. $B$ is an arbitrary functional of $g$; the gauge fixing functional
which gives the gauge fixing condition, while $\mathcal{F}$ is the variational derivative of the gauge fixing function with respect to the gauge group parameters. 
Raising and lowering of indices is with respect to the Minkowski flat metric Eq.~(\ref{future metric}) and  ``$i\epsilon$ terms'' 
are representative of the usual vacuum choice terms in the propagator \cite{Weinberg:1995mt, Weinberg:1996kr}. 
The gauge-fixing functionals are equivalent to introducing the usual ghost field Lagrangian in the Faddeev-Popov formalism. 
Using the co-vector transformation law, 
\begin{equation}
\label{SP2}
a_\mu=\sum_{\nu} a'_\nu 
\frac{\partial y^\nu}{\partial x^\mu}\, ,
\end{equation}
we can redefine the path integral in terms of retarded coordinates Eq.~(\ref{future co-ordinates}). The tensor transformation law is linear in the 
fields, therefore it yields a field independent factor. 
Thus changing the integration variables in the path integral does not change any of the correlation functions. 
Note, however, that in order to obtain a well formulated 
path integral, we require the Hermitian operators to constrain their eigenvalues to 
the set of real numbers, this technicality is why we delay using the usual $\{z,\bar{z}\}$ 
coordinates until the very end (although path integrals in both coordinate systems are equally valid of course). 
We will use the Faddeev-Popov prescription with the gauge condition Eq.~(\ref{gauge}). 
\begin{equation}
\label{SP3}
D\left[ \mathcal{J} \right]= \int\prod_{\nu,y}\left(dA_{\nu} \left(y \right) \right)\delta 
\left(A_u|_{\Dot{M}}\right)\delta \left( A_r \right) 
\exp \left[i\int_{\mathcal{M}}\left( -\frac{1}{4e^2} F_{\mu\nu} F^{\mu\nu} 
+ i\epsilon \, \mathrm{terms}\right)+i\int_{\partial\mathcal{M}}A_B\mathcal{J}^B \right] \, .
\end{equation}
In the above and in what follows, we will temporarily ignore the past null infinity boundary, 
but we will include it later as the analysis and the obtained result is completely analogous 
to the future null infinity case (apart from a sign flip). 
In the path integral, the gauge fixing functional is chosen so as to enforce Eq.~(\ref{gauge}).
In the above and what follows, the index $i$ runs over the components $\{ \theta, \phi \}$. 
The external current coupling is extremely useful, as it will allow us to find all desired 
boundary vacuum correlation functions with relative ease,
\begin{equation}
\label{effectivecurrent}
\left. 
\left( -\frac{\delta^2}{\delta\mathcal{J}_A\delta\mathcal{J}_B} \right) \right|_{\mathcal{J}=0}
D \left[\mathcal{J} \right] = \left<T \left\{ \left. A_A \right|_{\partial\mathcal{M}}\left. A_{B} \right|_{\partial\mathcal{M}} 
\right\}\right> \, ,
\end{equation}
where $T\{\cdots\}$ expresses the conventional time ordering. 
By dividing the Minkowski manifold into the boundary $\dot M$ and 
the interior $\mathfrak{int}\left[M\right]$ (all but the last slice in the discretization of 
the radial direction), or roughly speaking,
$\mathfrak{int}\left[M\right]$, 
\begin{align}
\label{Int}
D_{\mathfrak{int}}\left[ A_\nu|_{\Dot{M}} \right]\approx\int_{A_\nu|_{\Dot{M}}}
\prod_{\nu,y\in\mathfrak{int}[M]}\left(dA_{\nu} \left(y \right) \right)\delta 
\left( A_r \right)\exp \left[i\int_{M} dudrd^2z\sqrt{-g}\left( -\frac{1}{4e^2}
F_{\mu\nu} F^{\mu\nu}+i\epsilon \, \mathrm{terms} \right)
\right] \, .
\end{align}
Explicitly evaluating this path integral is non-trivial. Luckily, all we need to know and make use of is the fact that the integrals of quadratic Gaussian integrals 
(such as those of free quadratic theories when discretized) are also quadratic.  
Therefore, we infer that we can indeed write the path integral on the boundary as a quadratic ``action''. 
Additionally, the time-ordered correlation functions calculated from the quadratic action must 
be exactly equal to the ones calculated by other methods on the boundary. 
Moreover, the only fields that can possibly be included on the boundary are 
the $S^2$ components of the gauge field, as the other components are gauged out. 
This motivates the existence of a non-trivial quadratic effective action of boundary modes in the sense of Eq.~(\ref{effectivecurrent}), 
for as much as the time-ordered correlation functions themselves are well-behaved.
We will also use $\mathrm{SL}\left( 2, \mathbb{C} \right)$ invariance of 
celestial sphere amplitudes (from the Lorentz invariance at null infinity), 
\begin{equation}
\label{SL2C}
z\xrightarrow[]{SL(2,\mathbb{C})}\frac{az+b}{cz+d}\, , \quad \mbox{where} 
\quad ad-bc=1 \quad \mbox{and}\quad a,b,c,d\in\mathbb{C}\, ,
\end{equation}
Utilizing the above, we can fix the form of our effective boundary mode action, 
with external current coupling, up to scalings of the boundary gauge fields 
by demanding exact correspondence with the Ward identities and/or 
the operator product expansions (OPEs) of the Celestial Correlator formalism 
Eq.~(\ref{Soft}), as this just a different way of deriving the same results. 
This would spare us the explicit calculation of path integrating over the gauge modes 
on the interior of our space-time. 

Alternatively, we motivate the use of the boundary terms appearing in the Lagrangian formalism as the quadratic effective action 
appearing in the effective boundary path integral. 
To evaluate these boundary terms, we minimize the action as to satisfy the Euler-Lagrange E.o.M,
\begin{align}
\label{minaction}
S_{\mathrm{U(1)}}^{\mathrm{min}}&\dot{=}\int_{\mathcal{M}} 
\left[ -\frac{1}{4e^2} \bar{F}_{\mu\nu}\bar{F}^{\mu\nu}+ i\epsilon \, \mathrm{terms} \right] 
+\int_{\partial\mathcal{M}}\bar{A}_B\mathcal{J}^B \, , \nonumber \\
&=\int_{\partial\mathcal{M}}dud^2z\sqrt{-\gamma}\left[\frac{1}{4e^2}\partial_u\left(A_BA^B\right) 
+\frac{1}{2}A_B\mathcal{J}^B +i\epsilon \, \mathrm{terms}\right] \, .
\end{align}
When path-integrating over quadratic field theories, the path-integral fixes the integration variables to the value 
that minimizes the arguments of the exponential. 
In continuous field theories, this is often argued to mean the values of the fields that satisfy the Euler-Lagrange E.o.M. 
However, one may ask about what happens at the boundary of the integration region, for example at $\partial\mathcal{M}$. 
The value of these fields need not be fixed by the E.o.M, as they may be taken as the Cauchy initial data, like what is seen in Eq.~(\ref{cascade}). 
Therefore, if one integrates out the modes that are indeed fixed by the E.o.M. and the Cauchy data using the fixed point caveat, assuming 
that these integrals do not couple the boundary modes to one another via an effective bulk contribution 
(an assumption that is certainly non-trivial, and is in fact false in general), then the remaining effective path integral 
would be of the form obtained from minimizing the action Eq.~(\ref{minaction}),
\begin{equation}
\label{SP6}
D\left[\mathcal{J}\right]=C\int\prod_{ \nu\in{z,\bar{z},u},y\in{_{\partial\mathcal{M}}}} 
\left(dA_{\nu} \left(y \right) \right)\exp\left(i\int_{\partial\mathcal{M}}dud^2z 
\sqrt{-\gamma}\left[\frac{1}{4e^2}\partial_u\left(A_BA^B\right)+\frac{1}{2}A_B\mathcal{J}^B 
+ i\epsilon \, \mathrm{terms}\right]\right) \, .
\end{equation}
Additionally, we can perform the $u$-integral and study the theory on $S^2$ by setting,
\begin{equation*}
\mathcal{J}_B\left(u,z,\bar{z}\right)=\lim_{a\rightarrow\pm\infty}\delta\left(u-a\right)\mathcal{J}_B\left(z,\bar{z}\right)\, .
\end{equation*}
Following the literature, we can define ``the soft current'' $N_B$ and ``the Goldstone current'' $\Phi_B$, respectively 
as in \cite{Nande:2017dba},
\begin{align}
\label{Soft}
N_B\equiv A_B|_{u=\infty}-A_B|_{u=-\infty}\, , \quad 
\Phi_B\equiv A_B|_{u=\infty}+A_B|_{u=-\infty} \, ,
\end{align}
where once again we suppress the contribution by past null infinity. 
We also utilize the absent magnetic monopole condition \cite{Strominger:2017zoo}, 
\begin{align}
\label{nomm1}
e^2 \partial_B N\equiv N_B, \quad \partial_B\Phi&\equiv \frac{1}{2}\Phi_B \, .
\end{align}
Taking the variational derivatives with respect to the $S^2$ external currents, 
after integrating out the $u$ dependence in Eq.~(\ref{SP6}), we obtain 
(the replacement Eq.~(\ref{nomm1}) is done after evaluating the correlation 
functions of Eq.~(\ref{Soft})),
\begin{align}
\label{SoftSym}
\left<N\left(z,\bar{z}\right)N\left(w,\bar{w}\right)\right>&=0 \, , \\
\left<N_z\left(z,\bar{z}\right)\Phi\left(w,\bar{w}\right)\right>&
=\frac{e^2}{4\pi}\frac{i}{z-w} \, , \\
\left<N\left(z,\bar{z}\right)\Phi\left(w,\bar{w}\right)\right>&
=\frac{i}{4\pi}\ln{|z-w|^2} \, .
\end{align}
These may look familiar, as the R.H.S. are exactly the factors appearing in 
the operator product expansions of the formalism in \cite{Strominger:2017zoo,Nande:2017dba}. 
However, a key difference with Ref.~\cite{Nande:2017dba} 
is the time ordered insertions of two soft Goldstone currents, $\Phi_i$,
\begin{equation*}
\left<\Phi_z\left(z,\bar{z}\right)\Phi_w\left(w,\bar{w}\right)\right>
= \frac{k}{\left(z-w\right)^2}  \, . 
\end{equation*}
While our ansatz of Eq.~(\ref{minaction}) dictates that this should be 
trivial $(k=0)$, there appears a non-trivial Kac-Moody level 
$(k=\Gamma_{\mathrm{cusp}}=\frac{e^2}{4\pi^2})$ in \cite{Nande:2017dba}. 
This is due to the fact that we are studying free Maxwell theory, and not massive (or massless) QED as is the case with the references.
Of course, one may add an $N_BN^B$ term  to the effective action to fix 
this result by hand, but this addition may seem unnatural. 
To see this, note that if one demands local interactions (an assumption that certainly would not hold in general as we are considering an effective theory on the boundary), 
$\mathrm{SL}(2,\mathbb{C})$ invariance and works in the vanishing monopole 
sector, Eq.~(\ref{nomm1}), then the existence of a quadratic boundary action 
forces the above correlation function equal to that of the first equation 
in Eq.~(\ref{SoftSym}), in other words, $k=0$. 
Furthermore, the $k$ is pure real, and thus must come from an imaginary contribution to the QED action. 
Therefore, the assumptions that we made thus far were far too simple and therefore yielded a trivial Kac-Moody level. 
However, an important hint provided by the above calculations is that the boundary terms of the action DO play 
a very important role in describing the dynamics of soft modes, as can be seen from the fact that the correlation 
functions obtained are of the correct form, even with the above set of naive assumptions. 
In fact, we will see that effectively when one path-integrates over all fields but $N$ and $S$ that a term proportional 
to $N_BN^B$ is indeed the only permitted term.

\subsection{Soft Effective Action as an IR Counter-term}

As a reminder of the previous section, we saw that the large-gauge residual symmetries of QED act on the action non-trivially; 
they modify the action by a boundary term. 
This effectively ``breaks'' (in the literal sense of spontaneous symmetry breaking if one works with the assumptions of subsection \ref{ad}) 
the large-gauge symmetry as the configurations in the same large-gauge class are no longer on equal footing, and the variational procedure is ailed. 
To remedy this, we proposed the enforcement of a boundary condition Eq.~(\ref{bc}). 
One way to impose the vanishing of Eq.~(\ref{bc}) is to amend that form with an action that is constructed from the independent soft current 
and soft Goldstone modes; $\Phi$. 
The Goldstone mode will act as a Lagrange multiplier that enforces Eq.~(\ref{bc}) once integrated out. 
The Goldstone modes transform under the large-gauge transformations as,
\begin{equation}
\label{goldstone}
\Phi\left(z,\Bar{z}\right)\rightarrow \Phi\left(z,\Bar{z}\right)+\epsilon\left(z,\Bar{z}\right) \, ,
\end{equation}
with the amended action transforming as,
\begin{equation}
\label{transf}
\delta_\epsilon S_{\mathrm{IR}}=\frac{1}{e^2}\int_{S^2} d^2z \sqrt{\gamma}\,\{\gamma^{AB}\nabla_A\epsilon\left(z,\Bar{z}\right)N_B\} \, ,
\end{equation}
then we immediately find that the soft Goldstone mode couples to the soft current mode, and acts as Lagrange multiplier. 
Therefore, up to an constant that can be fixed using the cusp anomalous dimension, we have the most general quadratic IR counter-term action (using the condition Eq.~(\ref{nomm1})),
\begin{equation}
\label{IReffective}
S_{\mathrm{IR}}\propto\int_{S^2} d^2z \sqrt{\gamma}\,\gamma^{AB}\{\nabla_A\Phi\nabla_BN+ae^2\nabla_AN\nabla_BN\} \, .
\end{equation}
This is, up to a factor of $2$ , which is immaterial as this vanishes identically once one integrates out the Goldstone mode, 
the form of the effective action obtained in \cite{Kalyanapuram:2021bvf}. 
With the correct proportionality constant, the relevant correlation functions are,
\begin{align}
\left<\Phi\left(z,\Bar{z}\right)N\left(w,\Bar{w}\right)\right>=&\frac{1}{4\pi}\ln\left[\left(z-w\right)\left(\bar{z}-\bar{w}\right)\right] \, , \nonumber \\
\left<\Phi\left(z,\Bar{z}\right)\Phi\left(w,\Bar{w}\right)\right>=& - \frac{a'e^2}{\pi}\ln\left[\left(z-w\right)\left(\bar{z}-\bar{w}\right)\right] \, , \nonumber \\
\left<N\left(z,\Bar{z}\right)N\left(w,\Bar{w}\right)\right>=&0 \, .
\end{align}
Perhaps unsurprisingly, these are the expected Soft correlation functions with ``$a'$'' satisfying,
\begin{equation}
\label{aprime}
a'=\frac{1}{8\pi}\ln\left(\frac{\lambda_{IR}}{\Lambda_{UV}}\right)\,.
\end{equation}
We should note that the form of the action has been deduced by means of inverting the above correlation functions in \cite{Kalyanapuram:2021bvf}. 
Using this IR counter-term in the path integral formulation along with vertex operators encoding the soft operators in the soft/hard decomposition 
of the $S$-matrix (or alternatively the Celestial Correlator formalism) gives the desired correlators. 
Indeed, one may note the similarity with the Coulomb gas formalism and thus the charge conservation follows if one is to take 
the Soft factors to be vertex functions of the Goldstone mode via the neutrality condition \cite{DiFrancesco:1997nk}. 
We now turn to a calculation of the central charge, which classifies the relevant $2D$ CFT.

\subsection{Stress Energy Tensor}

The theory with combined fields, $N\left(z,\bar{z}\right), \phi\left(z,\bar{z}\right)$ is a $2D$ CFT, with both fields being dimensionless scalars. 
The action that we will explicitly use from here-on will be,
\begin{equation}
\label{normalizedaction}
S_{\mathrm{IR}}=2\int_{S^2}d^2z\sqrt{-\gamma}\left[\nabla_B\phi\nabla^BN+a'e^2\nabla_BN\nabla^BN\right]\,,
\end{equation}
The IR effective action's unregulated holomorphic stress-energy tensor component is given by,
\begin{equation}
\label{stressenergy}
\hat{T}_{zz}=\frac{2}{\sqrt{\gamma}}\frac{\delta S_{\mathrm{IR}}}{\delta\gamma^{zz}}=4\left[\partial_z\phi\partial_zN+a'e^2\partial_zN\partial_zN\right]\,.
\end{equation}
After regulation (subtracting the vacuum expectation value), one obtains the following correlation function for the regulated holomorphic stress-energy tensors,
\begin{equation}
\left<T_{zz}\left(z\right)T_{ww}\left(w\right)\right>=\frac{1}{\pi^2}\frac{1}{\left(z-w\right)^4}\,.
\end{equation}
For the quantum holomorphic counterpart defined as,
\begin{equation}
\label{stressquantum}
T\left(z\right)=-2\pi T_{zz} \, ,
\end{equation}
we have the following leading OPE,
\begin{equation}
\label{OPEstress}
T\left(z\right)T\left(w\right)\sim \frac{4}{\left(z-w\right)^4}+ O\left(\left(z-w\right)^{-2}\right)\, .
\end{equation}
Therefore, we can read off the central charge; c, of the Virasoro algebra for the effective soft $2D$ CFT as,
\begin{equation}
\label{central}
c=8\,.
\end{equation}

\section{Remarks and Conclusion}

To summarize the results, we have shown that the large-gauge transformations (residual gauge modes which do not die-off fast enough) do not leave the minimized action invariant. 
Instead they generate a term proportional to the unit-sphere Laplacian of the Soft current. 
This term has been already shown to generate the large-gauge transformations or equivalently the Soft Theorems \cite{Strominger:2017zoo}. 
The minimization of the action in the path-integral formalism subject to boundary conditions, as well as the well-posedness of the variational principle 
thus suggested that we introduce an IR counter-term, with the soft Goldstone modes playing the role of the Lagrange multiplier. 
This in turn provided an effective action of the Soft modes that is dual to a bosonic $2D$ CFT, whose central charge is evaluated. 
The Soft Goldstone currents define a Kac-Moody algebra, whose level depends on the the theory in question. 

Fortunately, a similar discussion also holds for gravity. 
Indeed one can find an effective Soft action by using the expected correlation functions of the soft graviton Goldstone modes \cite{Kalyanapuram:2021bvf}. 
Alternatively, one may find the effective action of gravity as a necessary counter-term needed to cancel logarithmic divergences \cite{Nguyen:2021qkt}. 
Going a step further, it is also noted that the super-rotation modes, which generate the sub-leading graviton theorem, also admit an effective action prescription 
given by the Alekseev-Shatashvili type action \cite{Nguyen:2020hot}.

Despite great advances in the studies of the Soft theorems, interesting questions still linger. 
For example, having shown that the leading soft modes admit an effective action prescription, is it possible (perhaps cascadingly) to generalize 
the statement for sub-leading and sub-sub-leading Soft theorems? 
Additionally, the residual gauge transformations seem to mix up physical and un-physical aspects of the gauge theories. 
It is by definition that pure gauge degrees of freedom are redundancies in the description that should not have any physical manifestation, 
and yet the same seems to not apply for large-gauge modes, in an apparent contradiction. 
Closely related is the question of whether the Hilbert space of our theory is properly defined, as the vacuum and other states transform 
non-trivially under the large-gauge transformation. 
Finally, one may wonder whether the BRST formalism holds the answer to this question, in a similar manner in which it does for generic local gauge theories.

\begin{acknowledgments}
This work is supported by the JSPS Grant-in-Aid for Scientific Research (C)
No. 18K03615 (S.N.).
\end{acknowledgments}

\appendix

\section{Spontaneous Symmetry Breaking in Finite Volume \label{A1}}

We consider the following action in two dimensions, 
\begin{equation}
\label{P1}
S= \frac{\phi_0^2}{2} \int dt dx \left( \dot \xi^* \dot \xi - {\xi^*}' \xi' \right) \, .
\end{equation}
Here $\dot\xi=\frac{\partial\xi}{\partial t}$ and $\xi'=\frac{\partial\xi}{\partial x}$ 
and $\phi_0$ is a constant. 
By defining a new field $\xi=\e^{i\frac{\phi}{\phi_0}}$ with a constant $\phi_0$, 
the action~(\ref{P1}) can be rewritten as the action of free massless real scalar field, 
\begin{equation}
\label{P2}
S= \frac{1}{2} \int dt dx \left( {\dot \phi}^2 - {\phi'}^2 \right) \, .
\end{equation} 
The action~(\ref{P1}) or (\ref{P2}) is invariant under the transformation of the $U(1)$ 
symmetry with a parmeter $c$, 
\begin{equation}
\label{P3}
\phi \to \phi + c \, .
\end{equation}
We also identify $\phi$ and $\phi + 2\pi \phi_0$ because $\xi=\e^{i\frac{\phi}{\phi_0}}$. 
The Noether charge corresponding to the symmetry transformation~(\ref{P3}) is given by 
\begin{equation}
\label{P4}
P\equiv \int dx \dot \phi \, .
\end{equation}
In fact, we find 
\begin{equation}
\label{P5}
\left[ ic P, \phi \left(x, t \right) \right] 
= ic \int dx' \left[ \dot\phi \left( x', t \right), \phi \left(x, t\right) \right] 
= ic \int dx' \left( - i \delta \left( x' - x \right) \right)
= c \, .
\end{equation}
Here we have used the canonical commutation relation, 
\begin{equation}
\label{P6}
\left[ \dot\phi \left( x', t \right), \phi \left(x, t\right) \right] 
=  - i \delta \left( x' - x \right) \, .
\end{equation}
Eq.~(\ref{P5}) tells that the symmetry corresponding to the transformation~(\ref{P3}) 
is spontaneously broken. 
Let denote the vacuum state by $\left| 0 \right>$. 
If the symmetry is not broken, we find $P\left| 0 \right>=0$. 
By using \ref{P5}), however, we find 
\begin{equation}
\label{P6A}
\left< 0 \right| \left[ P, \phi \left(x, t \right) \right] \left| 0 \right> 
= -i \, ,
\end{equation}
which tells $P\left| 0 \right>\neq 0$ and therefore the symmetry 
orresponding to the transformation~(\ref{P3}) is spontaneously broken. 
We should note that $\phi$ is nothing but the Nambu-Goldstone boson. 

If the volume of space is infinite, the operator $P$ is ill-defined. 
We may define another operator $Q$ by 
\begin{equation}
\label{P7}
Q \equiv \int dx \phi \, .
\end{equation}
We might expect that $Q$ is the operator to shift the eigenvalue of $P$, or 
$P$ is the operator to shift the eigenvalue of $Q$. 
We find, however, the commutation relation between $P$ and $Q$ is not well-defined, 
\begin{equation}
\label{P8}
\left[ P, Q \right] 
= \int dx \int dx' \left[ \dot\phi \left( x', t \right), \phi \left(x, t\right) \right] 
= \int dx \int dx' \left( - i \delta \left( x' - x \right) \right)
= -i \int dx \to \infty \, .
\end{equation}
This tells that there does not exist the operator to shift the 
eigenvalue of $P$ or $Q$ and therefore the Fock spaces with different eigenvalues of 
$P$ or $Q$ completely decouple with each other. 

On the other hand, if the volume of the space is finite, all operator is well defined. 
We assume $x$ has a periodicity of $L$, $x\sim x + L$. 
Then $P$ in (\ref{P4}) and $Q$ in (\ref{P5}) are written as 
\begin{equation}
\label{P9}
P\equiv \int_0^L dx \dot \phi \, , \quad Q \equiv \int_0^L dx \phi \, ,
\end{equation}
and instead of (\ref{P8}), we obtain 
\begin{equation}
\label{P10}
\left[ P, Q \right] 
= -i L \, .
\end{equation}
Therefore the algebra is well-defined. 
Even in case of the finite volume, Eq.~(\ref{P6A}) tells the $U(1)$ symmetry is 
spontaneously broken. 
Let a vacuum $\left| 0; q \right>$ by 
\begin{equation}
\label{P11}
Q \left| 0; q \right> = q \left| 0; q \right> \, .
\end{equation}
Here $q$ is the eigenvalue of $Q$. 
If we consider a new vacuum state $\left| 0; q + c\right>$ by 
\begin{equation}
\label{P12}
\left| 0; q + c\right> = \e^{i\frac{c}{L} P} \left| 0; q \right> \, ,
\end{equation}
we find 
\begin{equation}
\label{18}
Q \left| 0; q + c \right> = \left(q + c \right) \left| 0; q + c \right> \, .
\end{equation}
Therefore the vacuum becomes a non-trivial representation of $U(1)$ algebra. 

We may consider the Euclidean theory whose action is given by 
\begin{equation}
\label{17}
S= \int dz d\bar z \partial_z \phi \partial_{\bar z} \phi \, .
\end{equation} 
By Wick-rotating the radial direction as 
\begin{equation}
\label{P13}
z=\e^{i(t-x)}\, ,\quad \bar z =\e^{-i(t + x)} \, ,
\end{equation}
that is 
\begin{equation}
\label{P14}
t = - \frac{i}{2} \ln z\bar z \, , \quad 
x= \frac{i}{2} \ln \frac{z}{\bar z} \, ,
\end{equation}
we obtain 
\begin{align}
\label{P15}
& \partial_z = \frac{i\e^{- i(t-x)}}{2} \left( -  \partial_t + \partial_x \right) \, , \quad 
\partial_{\bar z} = \frac{i\e^{i(t+x)}}{2} \left( -  \partial_t - \partial_x \right) \, , \\
& dz = i \e^{i(t-x)} \left( dt - dx \right) \, , \quad 
d\bar z = - i \e^{-i(t+x)} \left( dt + dx \right) \, .
\end{align}
Therefore we can rewrite the action~(\ref{P12}) as follows, 
\begin{equation}
\label{P16}
S= \frac{1}{2} \int_{-\infty}^\infty dt \int_0^{2\pi} dx 
\left( {\dot \phi}^2 - {\phi'}^2 \right) \, ,
\end{equation} 
which is identical with the action~(\ref{P2}) in the finite volume with 
$L=2\pi$. 
Then the vacuum becomes a non-trivial representation of $U(1)$ algebra. 
We should also note that $\partial_\mu \phi$ generates the $U(1)$ Kac-Moody 
algebra.

\end{document}